%% file: ms.tex
\documentclass[runningheads]{llncs}
\pdfoutput=1
\usepackage{amsmath}
\usepackage{amsfonts}
\usepackage{xcolor}
\usepackage{graphicx}
\usepackage{fontawesome}
\usepackage{ifthen}
\usepackage{booktabs}
\usepackage{multirow}
\usepackage{subcaption}
\usepackage[referable]{threeparttablex}
\usepackage[ruled,vlined]{algorithm2e}
\usepackage{xcolor}
\usepackage{tikz}
\newboolean{doc_anonymous}   

\begin{document}
%
%
\title{Anomaly Detection on X-Rays Using Self-Supervised Aggregation Learning}
\setboolean{doc_anonymous}{false}   

\ifthenelse{\boolean{doc_anonymous}}
{
\author{Anonymous \textsuperscript{\faEnvelopeO}}
\authorrunning{Anonymous}
\institute{Anonymous Organization\\
\email{***@***.***}\\
\url{www.***.***}}
}

{
 \author{Behzad Bozorgtabar\inst{1, 2,3} \textsuperscript{\faEnvelopeO}
 \and Dwarikanath Mahapatra\inst{4} \and Guillaume Vray\inst{1} \and Jean-Philippe Thiran\inst{1, 2,3}}
 \authorrunning{B. Bozorgtabar et al.}
 \institute{Signal Processing Laboratory 5, EPFL, Lausanne, Switzerland \email{\{firstname.lastname\}@epfl.ch}
 \and {Department of Radiology, Lausanne University Hospital, Lausanne, Switzerland}
 \and {Center of Biomedical Imaging, Lausanne, Switzerland}
 \and  Inception Institute of Artificial Intelligence, Abu Dhabi, UAE
 }
 }

\maketitle              

\input{abstract.tex}

\input{introduction.tex}


\input{method.tex}

\input{results.tex}

\input{conclusion.tex}


\bibliographystyle{splncs04}
\bibliography{ms}

\clearpage
\input{appendix.tex}

\end{document}

%% file: abstract.tex
\begin{abstract}
Deep anomaly detection models using a supervised mode of learning usually work under a closed set assumption and suffer from overfitting to previously seen rare anomalies at training, which hinders their applicability in a real scenario. In addition, obtaining annotations for X-rays is very time consuming and requires extensive training of radiologists. Hence, training anomaly detection in a fully unsupervised or self-supervised fashion would be advantageous, allowing a significant reduction of time spent on the report by radiologists. In this paper, we present SALAD, an end-to-end deep self-supervised methodology for anomaly detection on X-Ray images. The proposed method is based on an optimization strategy in which a deep neural network is encouraged to represent prototypical local patterns of the normal data in the embedding space. During training, we record the prototypical patterns of normal training samples via a memory bank. Our anomaly score is then derived by measuring similarity to a weighted combination of normal prototypical patterns within a memory bank without using any anomalous patterns. We present extensive experiments on the challenging NIH Chest X-rays and MURA dataset, which indicate that our algorithm improves state-of-the-art methods by a wide margin.
\keywords{Anomaly detection \and X-rays \and Self-supervised learning \and Deep similarity metric.}
\end{abstract}

%% file: introduction.tex
\section{Introduction}
Currently, supervised based deep learning approaches are ubiquitous and achieve promising results for abnormality detection in X-ray images \cite{wang2017chestx}. However, many real-world datasets of radiographs often have long-tailed label distributions. On these datasets, deep neural networks have been found to perform poorly on rare classes of anomalies. This particularly has a pernicious effect on the deployed model if, at test time, we place more emphasis on minority classes of abnormal X-ray images. For example, in detecting rare lung opacities, e.g., such as pneumonia in chest X-rays (CXR), normal X-rays are much easier to acquire. Besides, examining radiographs and reporting work for the signs of abnormalities are very time consuming and require qualified radiologists.

Anomaly detection based methods \cite{baur2019fusing,alaverdyan2020regularized,gong2019memorizing,norlander2019latent} can be significantly useful in large-scale disease screening and spotting candidate regions for anomalies. Classical anomaly detection (AD) methods such as One-Class SVM (OC-SVM) \cite{scholkopf2001}, Local Outlier Factor \cite{breunig2000lof}, or Isolation Forest \cite{liu2008isolation} often fail to be effective on high-dimensional data or be scaled to large datasets. To alleviate this concern, state-of-the-art methods like Deep SVDD \cite{ruff2018deep} and Deep SAD \cite{ruff2019deep} consider learning deep CNN features as an alternative to classical one-class anomaly detection. However, the former suffers from the well-known problem of mode collapse, while the latter tends to ignore the underlying structure of the images as the pre-trained weights from autoencoder are sensitive to biased low-level features. Unsupervised learning based methods \cite{zhai2016deep,bozorgtabar2010comparison,bozorgtabar2011genetic,gong2019memorizing} empower us to exploit unlabeled data, thus can be considered as suitable approach for anomaly detection. Reconstruction-based methods \cite{zhai2016deep,zong2018deep,gong2019memorizing} use well-established convolutional autoencoders to compress and reconstruct single-class normal samples, but autoencoders can sometimes reconstruct abnormal samples well, yielding miss detection of anomalies at test time. New anomaly detection methods \cite{schlegl2017unsupervised,davletshina2020unsupervised,tang2019abnormal} built upon generative adversarial networks (GANs) \cite{bozorgtabar2019learn,bozorgtabar2019using,goodfellow2014generative} have shown promising anomaly detection performance by using GANs' ability to learn a manifold of normal samples. However, generated samples by GANs do not always lie at the boundary of real data distribution, which is necessary to distinguish normal images from abnormal ones. Recently, self-supervised methods \cite{golan2018deep,zhuang2019local,chen2020simple,he2020momentum} have been proposed to use unlabeled data in a task-agnostic way for extracting generalizable features, where the dataset can be labeled by exploiting the relations between different input samples, rather than requiring external labels. For example, self-supervised deep methods \cite{golan2018deep,gidaris2018unsupervised} proposed to train a classifier for which a self-labeled multi-class dataset is created by applying a set of geometric transformations to the images. However, these methods are domain-specific and cannot generalize over other data types.

\subsubsection{Contribution.} In this paper, we propose SALAD, short for \textbf{S}elf-supervised \textbf{A}ggregation \textbf{L}earning for \textbf{A}nomaly \textbf{D}etection on X-rays, a new training scheme that derives an aggregation learning from measuring the similarity between the estimated features of normal samples, to improve clustering and form prototypical patterns. We present a principled formulation to bypass tedious annotations and remove potential bias introduced by training. We show our method’s superiority to existing anomaly detection methods on X-ray datasets. We also highlight the limitations of the current state-of-the-art methods.


%% file: method.tex
\section{Method}
\begin{figure*}[t!]
\centering
\includegraphics[height=8.3cm, width=0.9\linewidth,trim=1 1 1 1,clip]{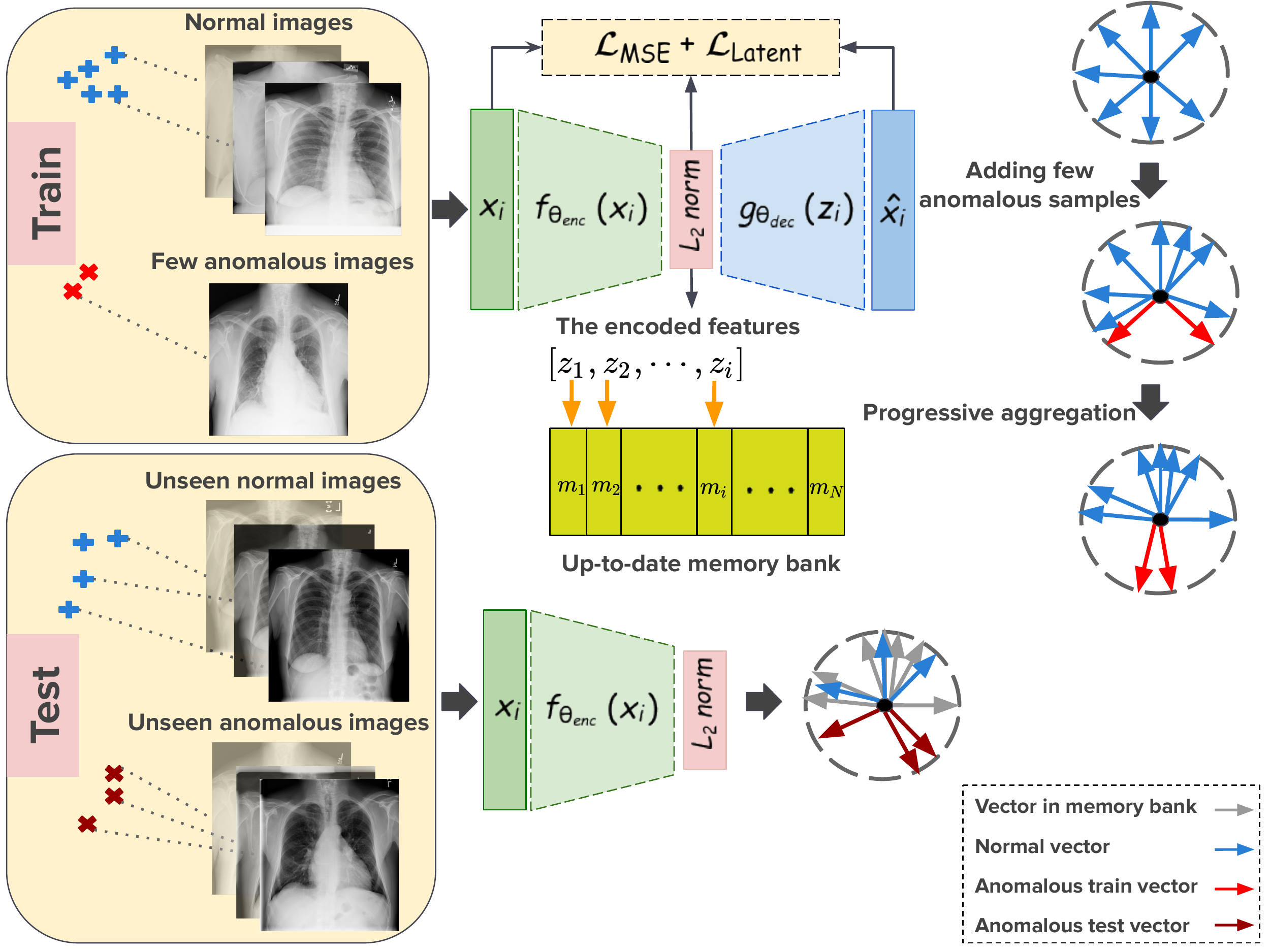}
\caption{\textbf{The SALAD pipeline.} The training process starts with forming the prototypical normal patterns (top). At test time, we measure the similarity between the test sample and normal patterns recorded in a memory bank (bottom).}
\label{fig:salad}
\end{figure*}

The merit of our approach is self-supervised representation learning, where the feature representation of each X-ray image is pushed closer to its similar neighbors, forming well-clustered features (prototypical patterns) in the latent space. This intuition is illustrated in Fig. \ref{fig:salad}. To do so, we propose to minimize the \textit{entropy} of each sample feature point's similarity distribution to other nearby samples. Learning feature similarity would require obtaining image embedding in the entire dataset. To avoid this, we use a memory bank \cite{wu2018unsupervised} to record and use the features. In every iteration, the memory bank is updated with the mini-batch features. The clustering objective will help us identify abnormal samples if they have different characteristics compared to normal prototypical patterns.

The backbone of our anomaly detection model is based on deep auto-encoder, where the encoder $f_{\theta_{enc}}: \mathcal{X} \rightarrow \mathcal{Z}$ is a convolutional neural network that represents input images $\left \{x_i \in \mathcal{X} \right \}^N_{i=1}$ in an informative latent domain $\mathcal{Z}$. The encoded representation performs as a query to compare with the relevant items in the feature memory bank. The decoder $g_{\theta_{dec}}: \mathcal{Z} \rightarrow \mathcal{X}$ is an up-sampling convolutional neural network that reconstructs the samples given their latent representations.

\subsubsection{Pre-Training.} For initialization, we establish an autoencoder pre-training routine using the image reconstruction loss (mean squared error), i.e. $\mathcal{L}_{\text{mse}}= \min\limits_{\theta _{enc}, \theta _{dec}} \left \| x- g_{\theta_{dec}} \circ f_{\theta_{enc}} \left ( x \right )\right \|_{2}^{2}$. In addition, we impose a constraint on the lower-dimensional representation of the data in which features of the same X-ray image under random data augmentations are invariant, while the features of different images are scattered. To do so, we jointly optimize the training of network with reconstruction loss and a sample specific loss $\mathcal{L}_{\text{ss}}$ \cite{wu2018unsupervised} to enforce a unique representation for each image:
\begin{equation}
\min\limits_{\theta _{enc}} \mathcal{L}_{\text{ss}} =  -\sum_{i \in \mathcal{B}_{\text{spl}}} \log{(\sum_{j \in AUG_i} p_{i,j})} \quad \text{s.t.} \quad p_{i,j} =\frac{\exp{(z_{j}^T z_{i}/\tau )}}{\sum_{k=1}^{N}\exp{(z_{k}^T z_{i}/\tau)}} 
\label{eq:1}
\end{equation}
where $\tau \in \left ( 0,1 \right ]$ denotes a fixed temperature hyperparameter. $\mathcal{B}_{\text{spl}}$ denotes the set of samples in the mini-batch. $z_{i}$ is the feature representation and $AUG_{i}$ denotes the set of randomly augmented versions of the image $x_{i}$. 

\subsubsection{Training.} 
\begin{figure*}[t!]
\centering
\includegraphics[height=7cm, width=0.8\linewidth,trim=1 1 1 1,clip]{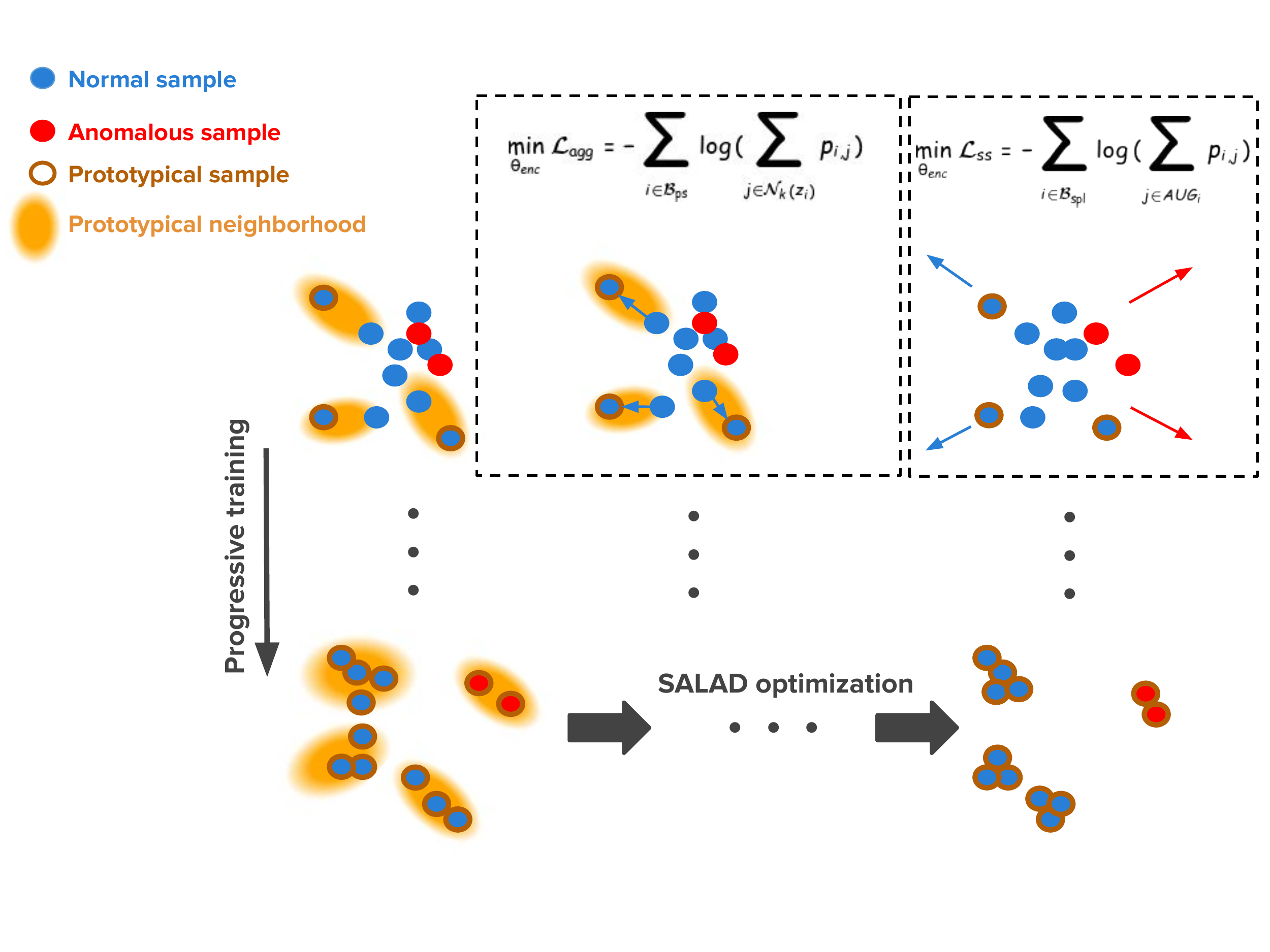}
\caption{\textbf{An overview of a proposed progressive training strategy.} We gradually increase sample neighborhoods to form prototypical patterns.}

\label{fig:progressive}
\end{figure*}

The learned feature representation at the pre-training stage may not preserve the similarity of different images. Therefore, we add the aggregation loss $\mathcal{L}_{\text{agg}}$ (Eq. \ref{eq:2}) to enforce consistency between samples lying in a neighborhood in latent space. We define aggregation loss as the (negative) log-likelihood that a specific sample will be identified as a member of the set of adjacent samples sharing the same prototypical pattern. This is achieved by the entropy measurement of the probability vector in Eq. \ref{eq:1}. The more similar the samples are, the less relative entropy they have. We progressively increase entropy to consider larger prototypical neighborhood for the samples and form clusters (see Fig. \ref{fig:progressive}). Finally, the proposed loss $\mathcal{L}_{\text{salad}}$ (Eq. \ref{eq:3}) joins all training losses:
\begin{equation}
    \min\limits_{\theta _{enc}} \mathcal{L}_{\text{agg}} = -\sum_{i \in \mathcal{B}_{\text{ps}}} \log{( \sum_{j \in  \mathcal{N}_k\left ( z_{i} \right )} p_{i,j})}
    \label{eq:2}
\end{equation}
\begin{equation}
    \min\limits_{\theta _{enc}, \theta _{dec}} \mathcal{L}_{\text{salad}} = \min\limits_{\theta _{enc}, \theta _{dec}} \mathcal{L}_{\text{mse}} + \lambda \min\limits_{\theta _{enc}} \overbrace{\left ( \mathcal{L}_{\text{ss}} + \mathcal{L}_{\text{agg}} \right )}^{\mathcal{L}_{\text{latent}}} 
    \label{eq:3}
\end{equation}
where $\mathcal{N}_{k} \left ( z_{i}  \right )$ denotes the top-k neighbours determined by the lowest cosine distance with respect to the embedding vector $z_{i}$. $\lambda$ is a hyperparameter to scale the losses ($\mathcal{L}_{\text{latent}}$) used in the latent space and $\mathcal{B}_{\text{ps}}$ denotes the set of prototypical samples in a mini-batch.



\subsubsection{Memory Bank.} 
Similarly to \cite{zhuang2019local,wu2018unsupervised}, we first initialize the memory bank with random unit vectors and then update its values $m_{i}$ using a weighted moving average scheme $m_{i}\leftarrow \left ( 1-t \right )m_{i}+tz_{i}$ considering the up-to-date features $z_{i}$, where $t$ is the fixed hyperparameter.

\subsubsection{Inference.} 
In the testing phase, an X-ray image is passed through the trained encoder and its representation is compared with the most relevant normal patterns in the memory bank for computing an anomaly score. Motivated by the weighted k-nearest neighbors (kNN), each vote $w_{i}$ is obtained from the top K nearest feature vectors in the memory bank. An anomaly score $\mathcal A\left ( \cdot  \right )$ is calculated by:

\begin{equation}\label{eq3}
\mathcal A\left ( x_{i} \right )=\frac{1}{K}\sum_{k=1}^{K}w_{i,k} \quad \text{s.t.} \quad   w_{i,k}=\frac{\arccos\left ( d\left ( z_{i},m_{k} \right ) \right )}{\sum_{j=1}^{N}\arccos\left ( d\left ( z_{i},m_{j} \right ) \right )}
\end{equation}
where $d\left ( \cdot ,\cdot  \right )$ denotes a cosine similarity, which computes similarity measurement between the test query feature $z_{i}=f_{\theta _{enc}}\left ( x_{i} \right )$ and the elements stored in the memory bank $\left \{m_{j} \right \}^N_{j=1}$. $\mathcal A\left ( x_{i} \right )$ is normalized to $\left [ 0,1 \right ]$. Ideally, the anomaly scores of anomalous images should be significantly larger than the scores from normal images. We also discard anomalous trained patterns in a memory bank as they can lead to adverse effects if anomalous prototypical patterns are similar to learned abnormal patterns.

%% file: results.tex
\section{Experimental Results}

\subsubsection{Datasets and Repartition.}
\begin{figure}[t!]
  \centering
  \includegraphics[width=1\textwidth]{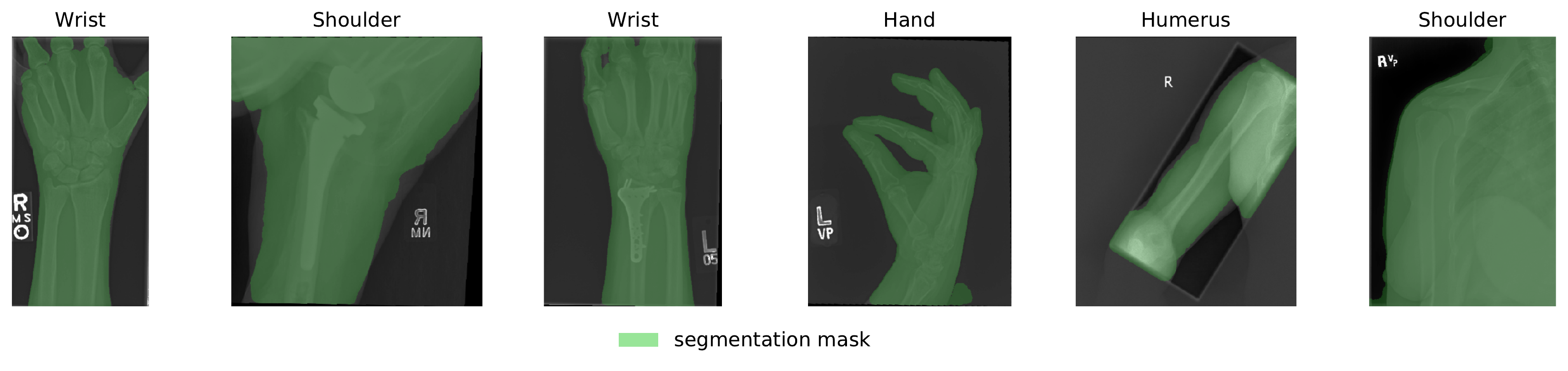}
  \caption{Examples of segmentation results of the musculoskeletal X-rays.}
  \label{fig:segmentation}
\end{figure}

\begin{figure}
  \centering
  \includegraphics[height=4.8cm, width=1\textwidth]{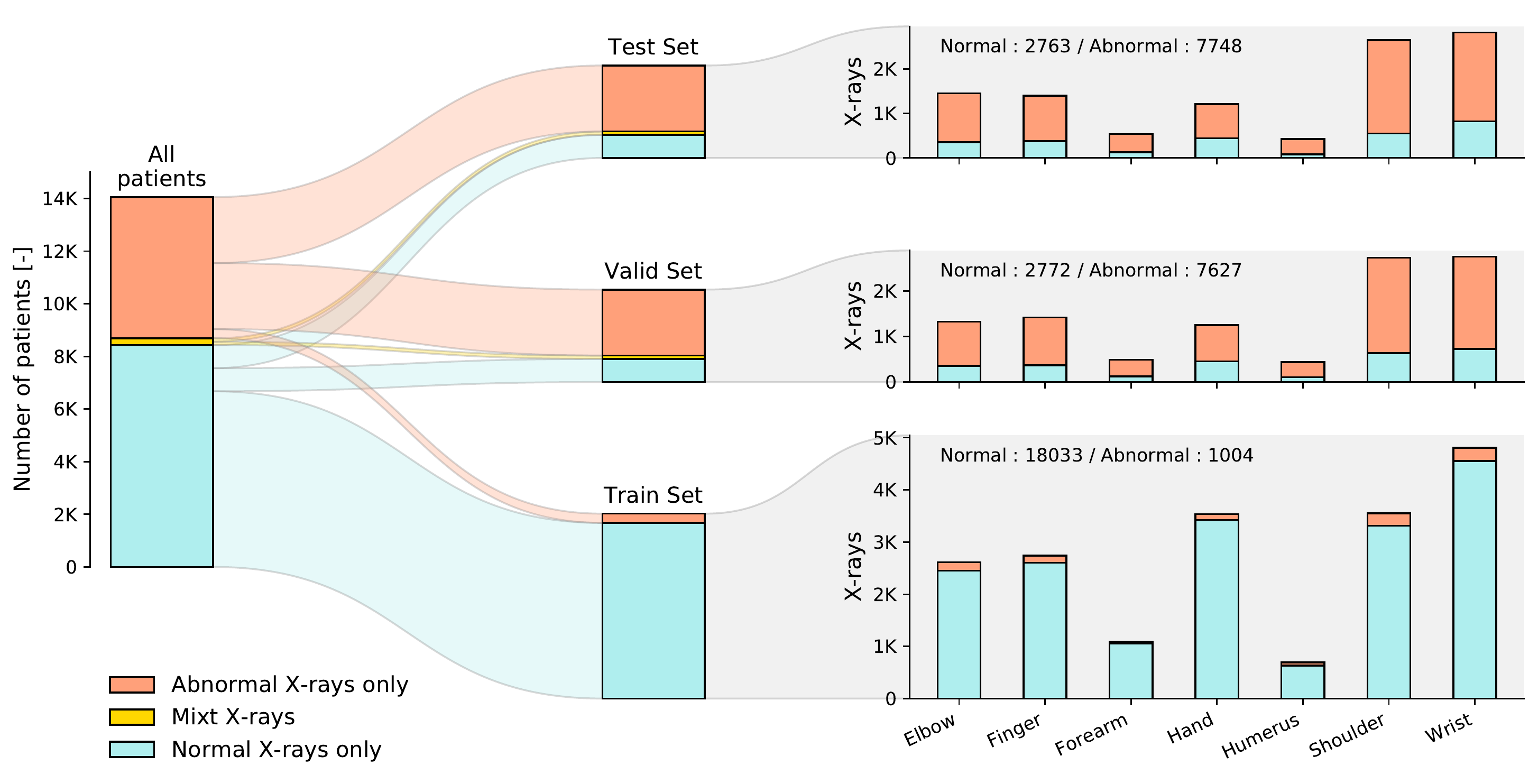}
  \caption{A visual summary of the applied data split scheme on the MURA dataset. }
  \label{fig:unsupervised_data_split}
\end{figure}

We validated our proposed method for classification of normal versus abnormal X-ray scans using two challenging public X-ray datasets, i.e., the NIH clinical center chest X-ray dataset \cite{wang2017chestx} and the MURA (musculoskeletal radiograph) dataset \cite{rajpurkar_mura_2018}. In NIH dataset \cite{wang2017chestx}, each radiographic image is assigned with diagnostic labels corresponding to 14 cardiothoracic or pulmonary diseases. We combine all CXRs with at least one of these 14 diseases into an aggregate abnormal class. For a fair comparison with \cite{tang2019abnormal}, we followed the same train, validation, and test subsets as in \cite{tang2019abnormal} so there was no patient ID overlap among the subsets. The MURA dataset \cite{rajpurkar_mura_2018} contains upper limb X-rays images labeled whether they contain anomaly or not. This dataset is composed of seven classes of body parts: finger, hand, wrist, forearm, elbow, humerus, and shoulder. There are a total of 40'005 X-ray images from 11'967 unique patients. We present a preprocessing pipeline, including the X-ray image carrier detection and unsupervised body part segmentation, using hysteresis thresholding by producing a binary mask (see Fig. \ref{fig:segmentation}). The splitting of the MURA dataset has been done based on the patient ID and the body part. This implies that all images of a specific body part from a given patient will be present in the same set. The patient's body parts are grouped into normal, abnormal, and mixed (meaning there are both normal and abnormal X-rays for that body part of a patient). The train set is composed mainly of normal samples (50\% of all the patient's body part and 95\% of normal samples) with few abnormal samples (5\%). The remaining normal, abnormal, and mixed samples are equally split between the validation and the test sets (see Fig. \ref{fig:unsupervised_data_split}).

\subsubsection{Implementation Details and Evaluation Metrics.} For the NIH dataset, we base our network architecture on the U-Net \cite{ronneberger2015u}, consisting of a 6-layer convolutional encoder network and a 6-layer up-sampling convolutional decoder network without skip connections (both have batch normalization and leaky ReLU after each layer). The last encoder output features are projected to a 200-dimensional space, and L2 normalized. We use Adam optimizer, $\left ( \beta _{1}=0.5,\beta _{2}=0.999 \right )$ and with a base learning rate of $0.0001$. We pre-train the network for 50 epochs. Then, we train the network progressively with $\mathcal{L}_{\text{salad}}$ in 10 rounds with 50 epochs per round. The images were resized to $256\times 256$ and we set $\tau =0.1$, $\lambda=0.25$, $t=0.5$ and $K=100$, respectively. The batch size was also set to 16. These optimum values are determined experimentally. We apply a slightly different experimental setup on the MURA dataset, and we replace the encoder and decoder with a ResNet-18 \cite{he2016deep} and a mirrored ResNet-18, respectively. Besides, the musculoskeletal X-ray images are resized and padded so that their major axis is 512 pixels long while keeping the aspect ratio. We adopt area under the ROC curve (AUC) and Area Under Precision-Recall Curve (AUPRC) as our evaluation metrics.




\subsubsection{Comparison with SOTA Unsupervised Methods.}
Fig. \ref{fig:salad_results} shows that our method significantly outperforms all recent anomaly detection methods, whether trained in an unsupervised mode, including OCGAN \cite{tang2019abnormal}, Deep SVDD \cite{ruff2018deep} and Deep Autoencoder (DAE) or with a self-supervised fashion (Geometric \cite{golan2018deep}). Although we use a few labeled anomalous data, the label information is not incorporated into our training, and we discard anomalous prototypical vectors for anomaly score calculation. Our method also achieves better performance compared to (GAN-GP) \cite{gulrajani2017improved}, where we replace GAN objective in \cite{tang2019abnormal}  with gradient penalty (Fig. \ref{fig:roc}).

\subsubsection{Comparison to Methods that Use a Small Pool of Labeled Anomalies.}
\begin{figure}[t!]
      \centering
      \includegraphics[width=1\textwidth]{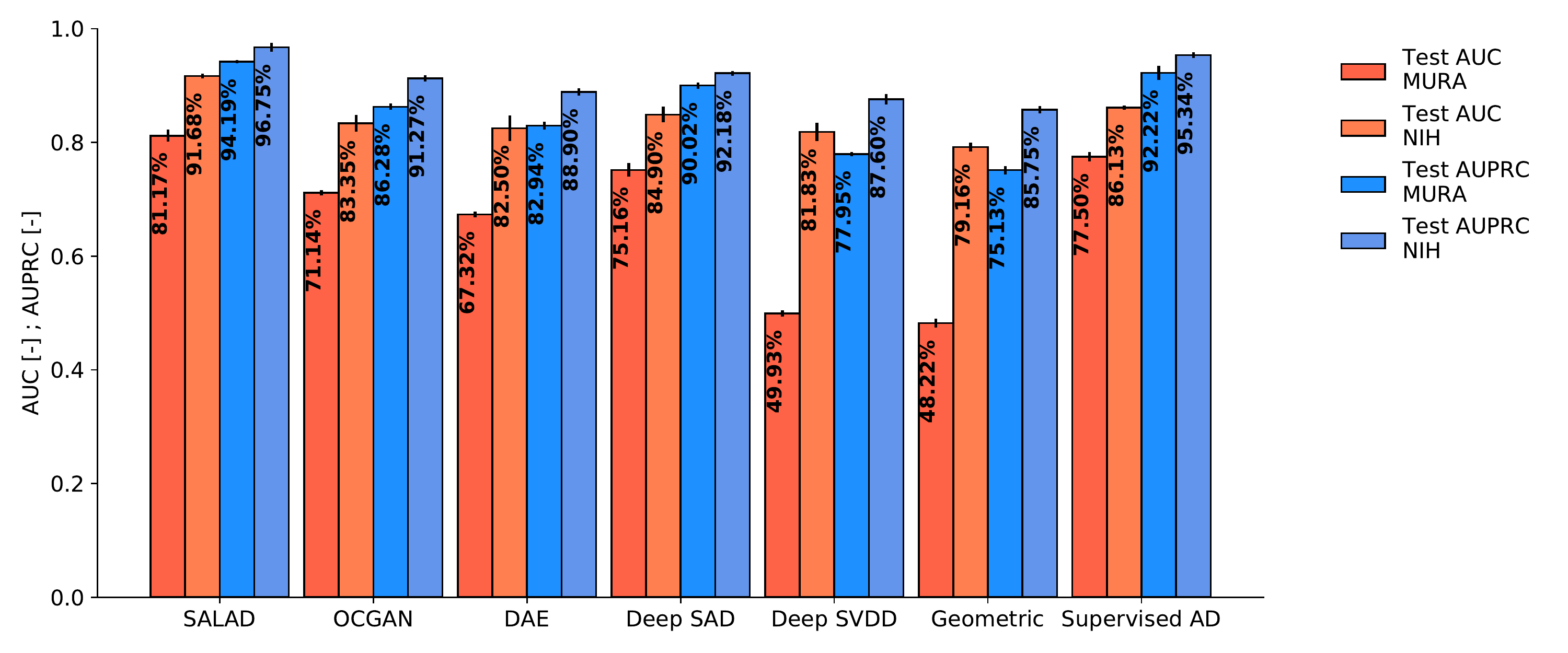}
      \caption{\textbf{AUC} and \textbf{AUPRC.} Comparison of different anomaly detection methods. The bar height represents the mean (AUC or AUPRC) over four replicates of training, while the error bar is a 95\% confidence interval computed as 1.96 std.}
      \label{fig:salad_results}
    \end{figure}

To establish competing methods, we compare our method with the state-of-the-art semi-supervised method, Deep SAD \cite{ruff2019deep}, with the same data splitting and network bottleneck as outlined above. In addition, we train the supervised classifier, ResNet-18 \cite{he2016deep} on the binary cross-entropy loss. We observed that our method surpasses other competing methods on two test sets (see Fig. \ref{fig:salad_results}). Semi-supervised and supervised approaches suffer from overfitting to previously seen anomalies at training while our self-supervised method generalizes well to unseen imbalanced anomalies (Fig. \ref{fig:statistics}).

\subsubsection{Ablations.}
\begin{figure}[t!]
\centering
\begin{subfigure}{.45\linewidth}
  \centering
  \includegraphics[height=4cm, width=\textwidth]{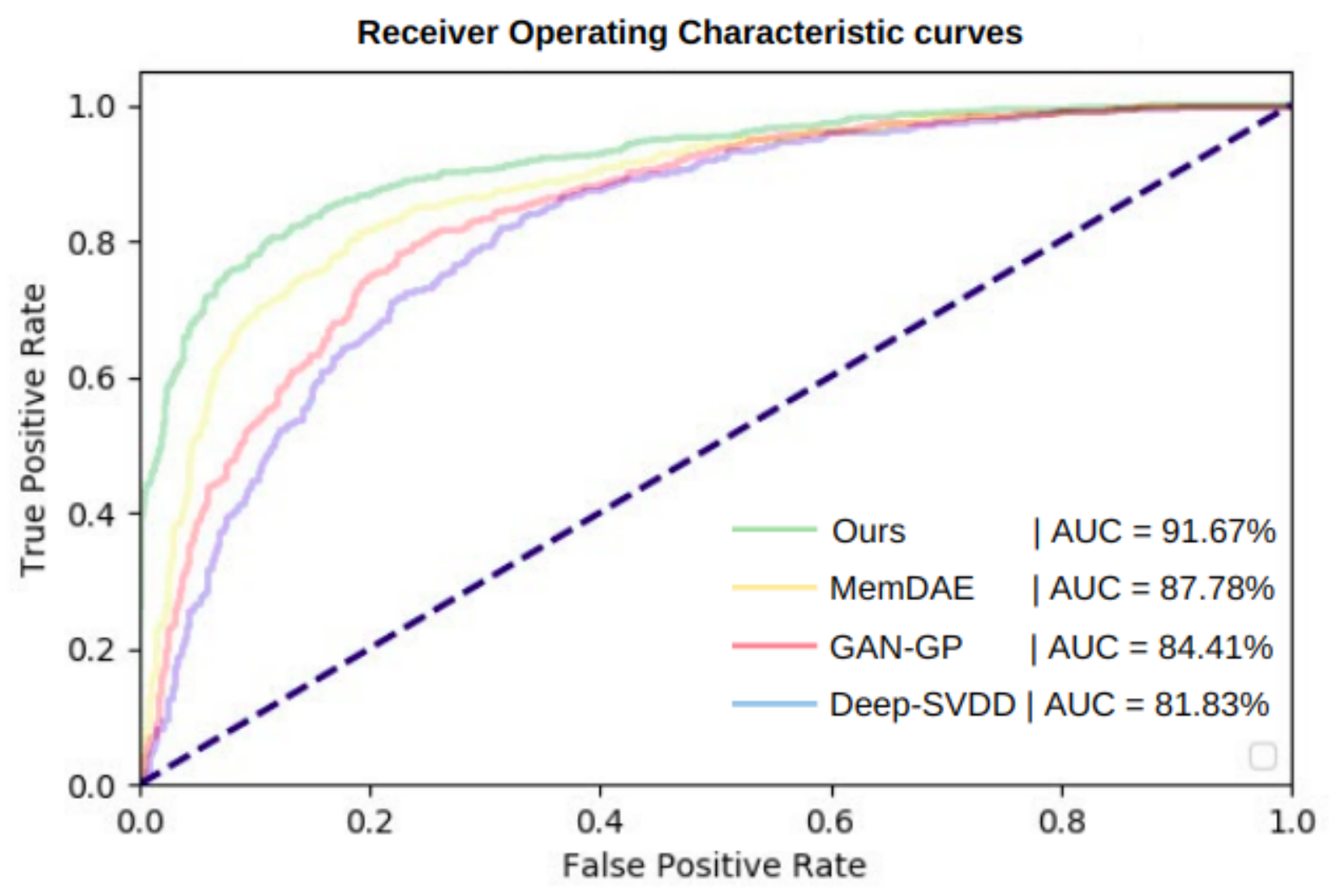}
  \caption{ROC curves }
  \label{fig:roc}
\end{subfigure}%
\begin{subfigure}{.45\linewidth}
  \centering
  \includegraphics[height=4cm, width=\textwidth]{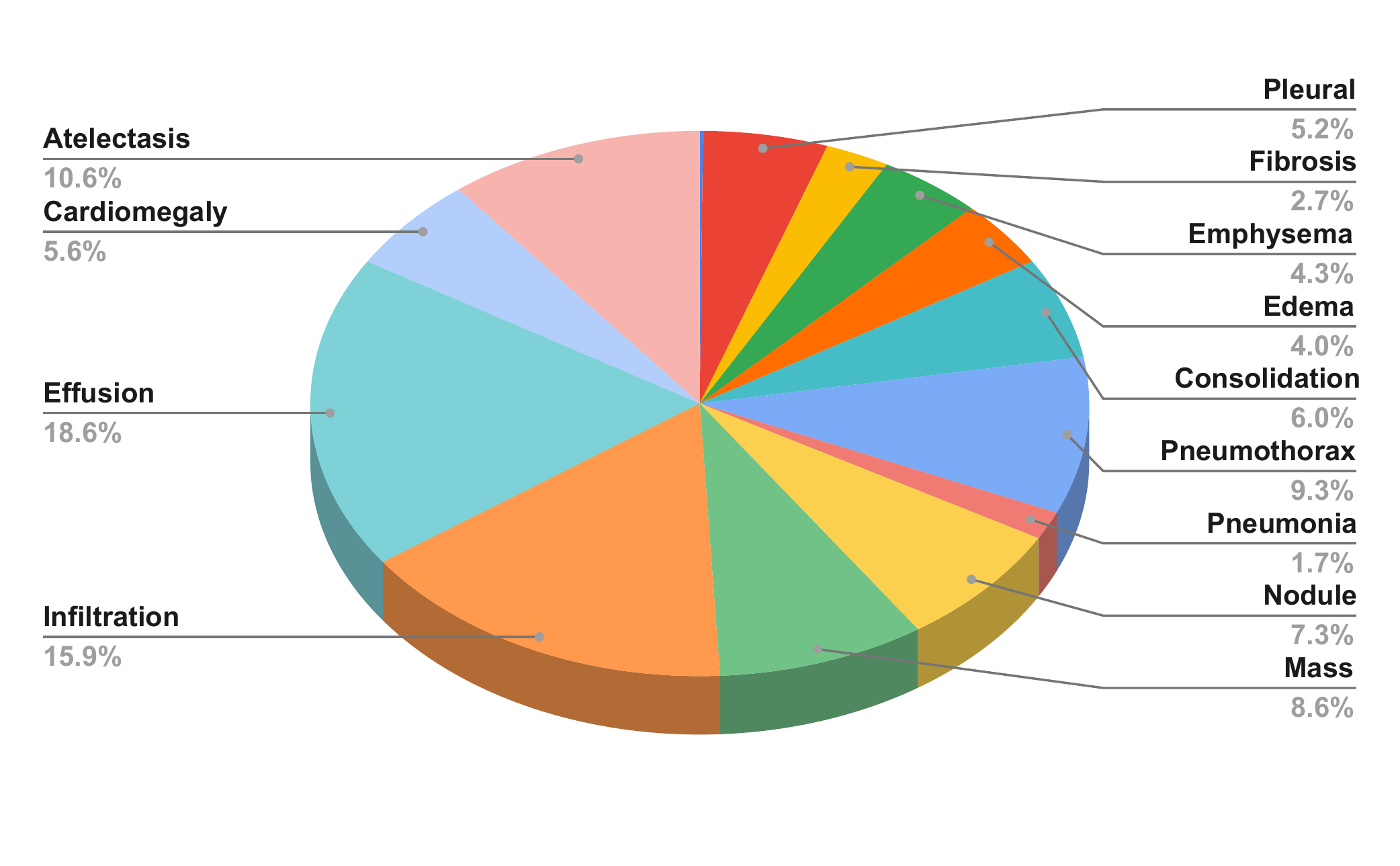}
  \caption{The statistics of the NIH set}
  \label{fig:statistics}
\end{subfigure}
\caption{(a) ROC curves comparison performances on the NIH dataset. (b) The NIH dataset statistics used in our experiments.}
\label{fig:ROC_statistics}
\end{figure}

We conduct a series of ablation studies to justify the effectiveness of our contributions by comparing our full model with the following alternatives, using: 1.  A Memory-based Deep Autoencoder (MemDAE) by turning off the proposed loss terms ($\mathcal{L}_{\text{agg}}$, $\mathcal{L}_{\text{ss}}$) and without using anomalous samples (Table 1 and Fig. \ref{fig:roc}); 2. Our method without the loss term $\mathcal{L}_{\text{agg}}$; 3. Our method without the loss term $\mathcal{L}_{\text{mse}}$. We observed that our method trained with each of the proposed loss terms, resulting in a notable performance gain over all the metrics, e.g., a gain of about 5.8\% in AUC, compared to MemDAE on the MURA dataset (Table 1).
Nevertheless, our baseline method (MemDAE) without anomalous samples, which is trained solely with MSE loss, outperforms all previous anomaly detection methods. We also conduct sensitivity analysis to investigate the effect of included labeled anomalies during training on final performance. To do so, we increased the ratio of known anomalous samples up to 15\% and observed that our method is not very sensitive to an anomalous ratio (Table 1). This can be explained by the fact that SALAD does not require label information. Instead, it uses anomalous samples to have a better separation of prototypical patterns.

\begin{figure}[t!]
  \centering
  \includegraphics[height=6.8cm, width=1\textwidth]{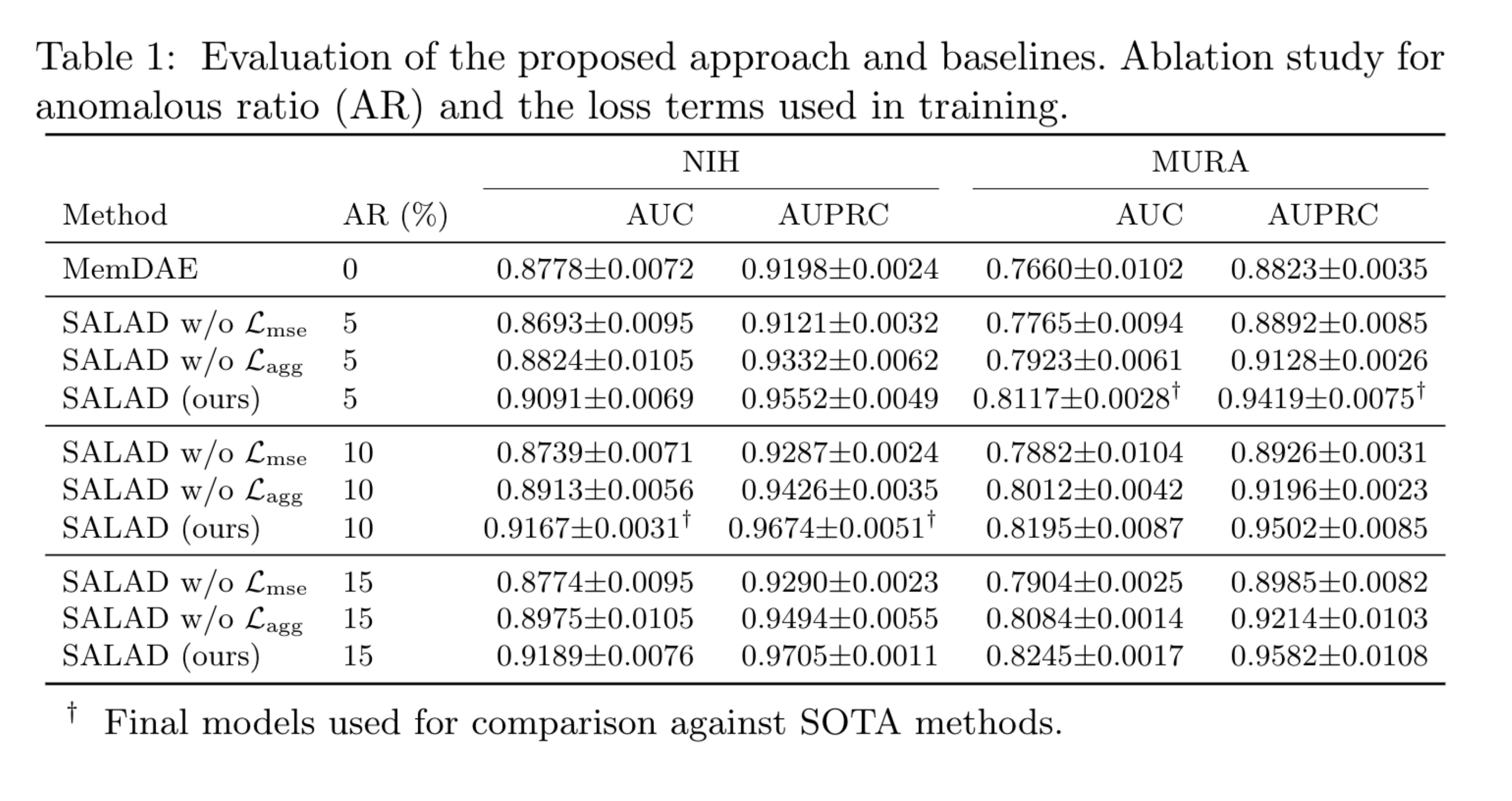}
  \label{fig:table_resultt}
\end{figure}

%% file: conclusion.tex
\section{Conclusion and Future Work}
In this work, we proposed SALAD, a self-supervised aggregation based learning framework for X-ray anomaly detection. This paper's novelty lies in jointly deep representation learning of X-ray images as well as aggregation criterion to distill out anomalous data. We use progressive training to enforce consistency between similar data samples in the embedding space to facilitate the formation of prototypical normal patterns. Hence, abnormal X-ray samples appear less likely to be represented by the normal learned patterns. SALAD achieves state-of-the-art anomaly detection results across all tested learning regimes, including unsupervised methods and those trained with small amounts of labeled data. As future work, we envision the broad application of our approach across different image modalities and beyond anomaly detection where the annotation is very costly, e.g., unsupervised domain adaptation.


%% file: appendix.tex
\section{Appendix}

\subsection{Data Repartition}

\begin{figure}[h]
      \centering
      \includegraphics[width=1\textwidth]{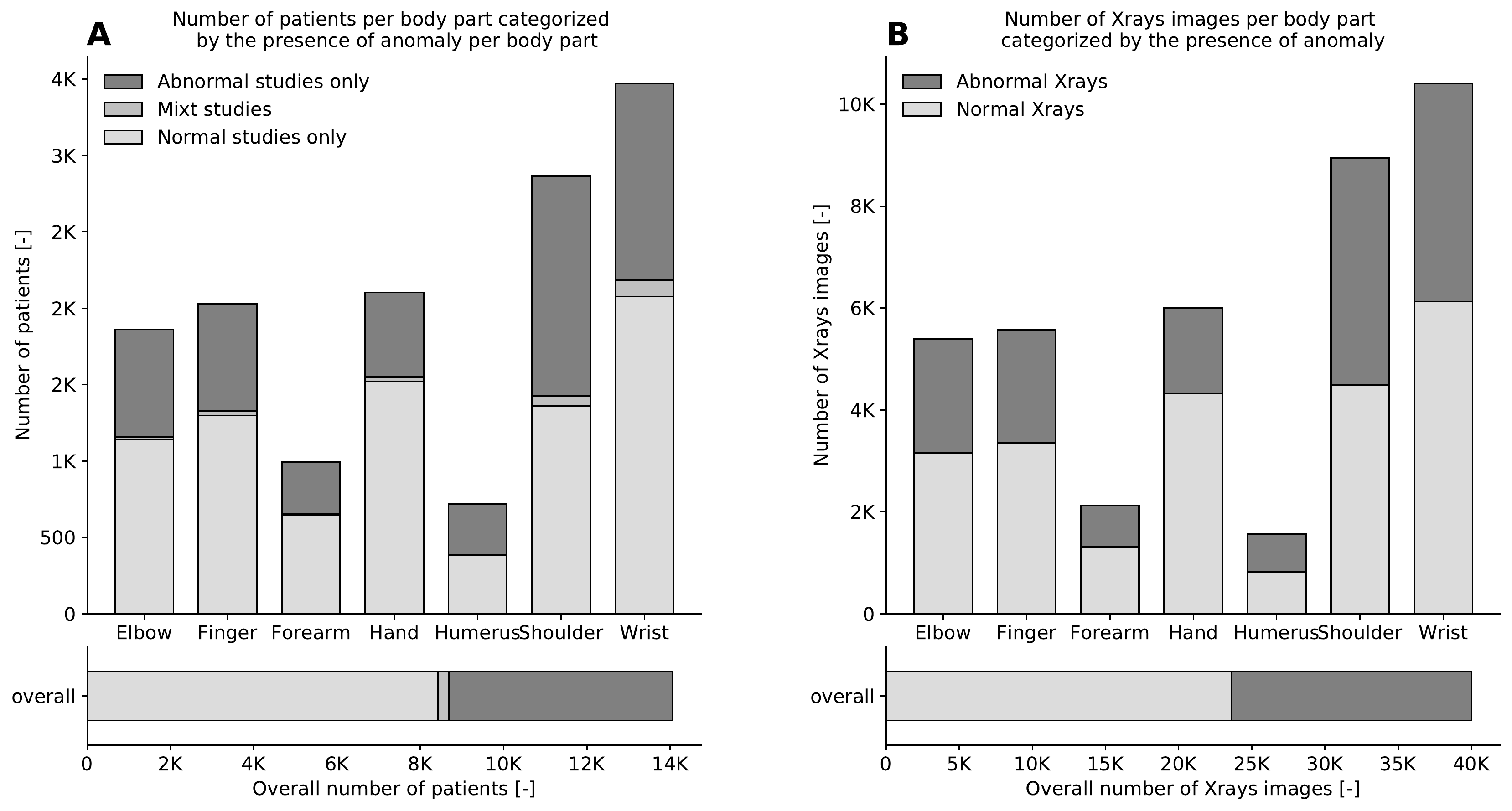}
      \caption{\textbf{Data repartition.} The two bar plots present the repartition of anomalies by body parts at two levels on the MURA dataset. (\textbf{A}) Repartition at the level of the patient's body part (each unit is a set of x-rays of a body part from one patient). Those patient's body parts are categorized into three groups: normal, abnormal, mixed (the patient has both normal and abnormal X-rays for the body part). (\textbf{B}) Repartition at the level of the x-ray images categorized as normal or abnormal.}
      \label{fig:data_repartition_new}
    \end{figure}
    
\begin{figure}[h]
      \centering
      \includegraphics[width=1\textwidth]{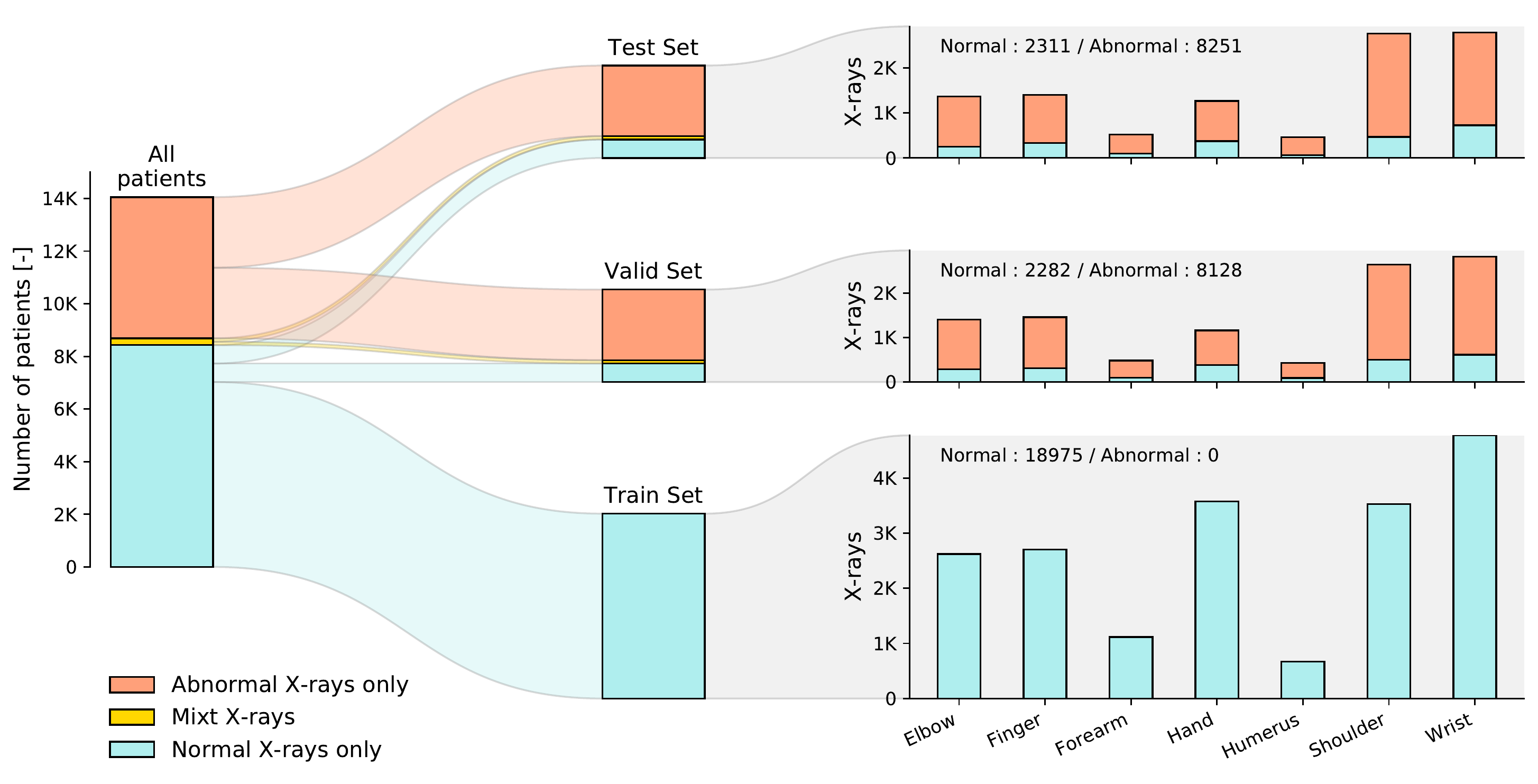}
      \caption{\textbf{Data splitting of the MURA dataset used for our unsupervised baseline (MemDAE ).} The unsupervised splitting at the level of the patient's body part is visually presented. The train set represents 50\% of all the patient's body part and is composed only of normal samples. On the left, the repartition of X-ray images by body part is shown.}
      \label{fig:unsupervised_data_split_new}
    \end{figure}
